\documentclass[pra,twocolumn,showpacs,amsfonts,amsmath,floatfix,superscriptaddress]{revtex4-1}

\usepackage[svgnames]{xcolor}

 \usepackage[
            colorlinks=true,        
            allcolors = black,  
            citecolor=DarkBlue, 
            linkcolor=DarkBlue,
            urlcolor=DarkBlue
        ]{hyperref}  
\usepackage{amsmath, amsfonts, amssymb, amsthm, bbm}
\usepackage{dsfont}
\usepackage{bm} 
\usepackage{mathtools}
\usepackage{times}
\usepackage{physics}
\usepackage{graphicx}
\usepackage{float}
\usepackage[colorinlistoftodos]{todonotes}
\usepackage{dcolumn}

\usepackage{diagbox}

\allowdisplaybreaks
\begin{document}
\title{Deterministic Gaussian conversion protocols for non-Gaussian single-mode resources}
\author{Oliver Hahn}%
\thanks{These authors contributed equally to this work.}
\affiliation{Department of Microtechnology and Nanoscience (MC2), Chalmers University of Technology, SE-412 96 G\"{o}teborg, Sweden}

\author{Patric Holmvall}
\thanks{These authors contributed equally to this work.}
\affiliation{Department of Microtechnology and Nanoscience (MC2), Chalmers University of Technology, SE-412 96 G\"{o}teborg, Sweden}
\affiliation{Department of Physics and Astronomy, Uppsala University, Box 516, S-751 20, Uppsala, Sweden}

\author{Pascal Stadler}
\affiliation{Department of Microtechnology and Nanoscience (MC2), Chalmers University of Technology, SE-412 96 G\"{o}teborg, Sweden}

\author{Giulia Ferrini}
\affiliation{Department of Microtechnology and Nanoscience (MC2), Chalmers University of Technology, SE-412 96 G\"{o}teborg, Sweden}

\author{Alessandro Ferraro}
\email{alessandro.ferraro@unimi.it}
\affiliation{Centre for Theoretical Atomic, Molecular and Optical Physics, Queen's University Belfast, Belfast BT7 1NN, United Kingdom}
\affiliation{Dipartimento di Fisica ``Aldo Pontremoli'',
Università degli Studi di Milano, I-20133 Milano, Italy}

\date{\today}

\begin{abstract}

In the context of quantum technologies over continuous variables, Gaussian states and operations are typically regarded as freely available, as they are relatively easily accessible experimentally. In contrast, the generation of non-Gaussian states, as well as the implementation of non-Gaussian operations, pose significant challenges. This divide has motivated the introduction of resource theories of non-Gaussianity. As for any resource theory, it is of practical relevance to identify free conversion protocols between resources, namely Gaussian conversion protocols between non-Gaussian states. Via systematic numerical investigations, we address the approximate conversion between experimentally relevant single-mode non-Gaussian states via arbitrary deterministic one-to-one mode Gaussian maps. First, we show that cat and binomial states are approximately equivalent for finite energy, while this equivalence was previously known only in the infinite-energy limit. Then we consider the generation of cat states from photon-added and photon-subtracted squeezed states, improving over known schemes by introducing additional squeezing operations. The numerical tools that we develop also allow to devise conversions of trisqueezed into cubic-phase states beyond previously reported performances. Finally, we identify various other conversions which instead are not viable.

\end{abstract}

\maketitle

\section{Introduction}

In the context of quantum information science, continuous-variable (CV) quantum systems~\cite{Weedbrook:2012tz, Serafini:2017uz} are constituted of indistinguishable bosons that can be prepared, manipulated, and measured in order to implement relevant information processing protocols. They stand at the forefront of quantum technologies and, more recently, they have gained prominence in the context of quantum computation \cite{grimsmo2021quantum, terhal2020towards} over a variety of physical platforms, such as optical~\cite{Pfister:2020vu} and microwave radiation~\cite{cai2021bosonic, ma2021quantum, Blais:2019tq, Grimsmo:2017tf,  Hillmann:2020uf}, trapped ions~\cite{Serafini:2009vq, Fluhmann:2019aa}, opto-mechanical systems~\cite{Schmidt:2012ww, Houhou:2015aa, Nielsen:2017vj, rakhubovsky2021stroboscopic, houhou2022unconditional}, atomic ensembles~\cite{Stasinska:2009ta, Milne:2012wz, Ikeda:2013wz, Motes:2017vd}, and hybrid systems~\cite{Aolita:2011ti}. 

A major feature of CV systems is their resilience to noise. In particular, their associated infinite dimensional Hilbert space can be exploited to host a variety of bosonic codes \cite{grimsmo2021quantum, terhal2020towards, cai2021bosonic, ma2021quantum} --- namely, sets of quantum states where logical digital information can be encoded redundantly to enable fault-tolerance against arbitrary errors.
In particular, the use of superconducting cavities in the microwave regime has allowed for reaching the break-even point for error correction~\cite{Ofek:2016wb}, meaning an enhancement in the lifetime of quantum information encoded in the state of the field using a rotationally symmetric bosonic code (RSB)~\cite{Grimsmo:2020wa} compared to an unencoded qubit using the same hardware. 

Quantum states over CV systems, and operations thereof, can be classified as Gaussian and non-Gaussian --- where the nomenclature stems from the corresponding Wigner functions~\cite{Ferraro:2005vh}. Such a divide emerges naturally from both practical and theoretical considerations: on the one hand, Gaussian states and operations are relatively easy to implement experimentally, in contrast to non-Gaussian ones; on the other hand, the Gaussian sector of the CV Hilbert space can be efficiently simulated on classical machines, whereas non-Gaussian components can enable universal quantum information processing~\cite{Lloyd:1999aa, Gottesman:2001aa} and even promote it to fault tolerance using non-Gaussian bosonic codes. This state of affairs has motivated the introduction and quantification of the concept of non-Gaussianity \cite{Genoni2007, Genoni2010, Marian2013a} and, more in general, the development of a resource theory of quantum non-Gaussianity~\cite{Albarelli:2018uu, Takagi:2018ul}. In other words, Gaussian states and operations are regarded as freely available whereas non-Gaussian states are promoted to the role of genuine resources.

As for any resource theory, the interconversion of resources plays a central role. For example, in the context of the resource theory of entanglement~\cite{Horodecki:2006vn}, the interconversion of entangled states using free operations (local operations and classical communication) is pivotal for quantum communication purposes, by means of protocols such as entanglement distillation. 
Since Gaussian operations are easily implementable and can hence be regarded as free operations, it is fundamental to
understand how different non-Gaussian states can be converted one to another by means of protocols that use only Gaussian operations~\cite{Albarelli:2018uu}. 
From a technological viewpoint, conversion protocols are important for applications to quantum computation as well as quantum communication~\cite{du2021conversion}. One context of application lies within the domain of distributed quantum computing architectures, where for example multiple superconducting quantum processors are connected by photonic links through the use of microwave optical converters. Different CV platforms suffer from different kinds of noise which can be, in turn, counteracted by means of different bosonic codes. It is therefore desirable to be able to freely (namely, using Gaussian protocols) convert between (non-Gaussian) bosonic codes.


In general, as much as desirable, a systematic study of Gaussian conversion protocols between genuinely quantum non-Gaussian states is elusive, due to the intrinsic difficulties stemming from the infinite dimension of the CV Hilbert space. Few attempts for specific cases have been considered in the literature. For example, in Ref.~\cite{yu2020} the Gaussian conversion of a trisqueezd state towards the cubic-phase state (see later for the specific definitions) has been analysed, and found that the fidelity between them could be significatively improved already by means of a deterministic Gaussian conversion protocol, described formally by a completely positive trace preserving (CPTP) Gaussian map. It is therefore tempting to study whether other meaningful deterministic Gaussian conversions between non-Gaussian states are possible to achieve with the sole use of Gaussian CPTP maps. 

In this work we perform a systematic numerical study of deterministic Gaussian conversions between single-mode non-Gaussian states, including bosonic code-words useful in quantum error correction, under the most general CPTP Gaussian map. To this aim, we implement the CPTP Gaussian maps in a numerically efficient way, taking full advantage of parallelism and high performance computing. This allows us to study conversions over a large scale of parameters, for various possible input and target states of the conversion protocols under consideration.

Motivated mainly by experimental attainability and applicative relevance, in the following we will consider a variety of non-Gaussian states: cat states, rotational symmetric bosonic codes, photon-added and -subtracted states, GKP codes, cubic-phase states and trisqueezed states. We will show that for most of these states it is hard to find conversions reaching high fidelities. However, we will also identify relevant cases in which excellent performances indeed can be achieved. In particular, we show that cat and binomial codes can be considered as approximately equivalent for a large set of parameters, beyond the known high-energy limit. Moreover we find that photon-added and photon-subtracted squeezed states, when acted upon with Gaussian conversion protocols, yield generation of cat states with significantly larger fidelities than what was previously shown. The numerical tools that we develop also allows for the conversion of trisqueezed into cubic-phase states beyond what was done in Ref.~\cite{yu2020}. More in general, beyond these specific examples, the numerical tools developed here can be used to test the interconvertibility between arbitrary single-mode non-Gaussian states under the action of deterministic CPTP maps.

The  paper  is  structured  as  follows.  In Section \ref{sec:background} we present the theoretical methods used to calculate bounds on the conversion fidelity, and define the states we are investigating, namely rotationally symmetric bosonic code-words, photon-added and photon-subtracted squeezed states, cubic-phase states, trisqueezed states and GKP-states. We also comment on the parameter ranges we used based on experimental implementations. In Section \ref{sec:result} we present the results for the different conversions that we addressed. Conclusive remarks are presented in Sec.~\ref{sec:conclusions}. Appendix \ref{app:numerics} briefly describes the numerical approach.

\section{Theoretical Background}
\label{sec:background}
In this work, we only consider conversions from one mode to one mode. Therefore we restrict all our definition to this case.
We are going to indicate the vector of quadrature operators as $\hat{\Vec{r}}=\qty(\hat{q},\hat{p})^T$. The quadrature operators are related to the  creation and annihilation operators by $\hat{q} = (\hat{a} +\hat{a}^\dagger)/\sqrt{2}$ and $\hat{p} =  ( \hat{a} - \hat{a}^\dagger)/(\sqrt{2}i)$, corresponding to setting $\hbar = 1$. 

Notable operations on the bosonic field that we will use in the following are implemented by the squeezing $\hat{S}(\xi)$, the displacement $\hat{D}(\beta)$ and the phase rotation $\hat{U}_p(\gamma)$ operators, which are defined respectively as
\begin{eqnarray}\label{eq:squeezing}
    \hat{S}(\xi)=e^{\frac{\xi^*}{2}\hat{a}^{2} -\frac{\xi}{2}\hat{a}^{\dagger 2}},
\end{eqnarray}
\begin{eqnarray}\label{eq:displacement}
    \hat{D}(\beta)=e^{{\beta\hat{a}^{\dagger }}-\beta^{\ast}\hat{a}},
\end{eqnarray}
\begin{eqnarray}\label{eq:phase-rot}
    \hat{U}_p(\gamma)=e^{-i\gamma\hat{n}},
\end{eqnarray}
with $\hat{n} = \hat a^{\dagger} \hat a$ the number operator, $\gamma \in \mathbb{R},\; \beta \in \mathbb{C},\; \xi \in \mathbb{C}$.
The subsequent application of squeezing and displacement to the vacuum state yields the squeezed coherent state expressed as
\begin{align}
\label{eq:squeezed-coherent}
  \ket{\alpha, r,\phi}=\hat{D}(\alpha) \hat{S}(r e^{-2i\phi}) \ket{0},
\end{align}
with $\alpha \in \mathbb{C},\; r \in \mathbb{R},\; \phi \in [0,2\pi)$.

\subsection{Characteristic Functions and CPTP Maps}
\label{sec:cptp_maps}

Completely-positive trace-preserving (CPTP) maps are called Gaussian if they map Gaussian states into Gaussian states
~\cite{Serafini:2017uz}.  These maps can be  characterized by their action onto the symmetrically ordered characteristic function
\begin{align}
\label{eq:generating-function}
    \chi_{\hat{\rho}}(\Vec{r}) = \Tr{ \hat{D}(-\Vec{r}) \hat{\rho}},
\end{align}
where $\hat{D}(-\Vec{r})$ is  the displacement operator of Eq.~(\ref{eq:displacement}), given by
\begin{align}
    \hat{D}(-\Vec{r}) = e^{-i( \Vec{r}^T \Omega \hat{\Vec{r}})}, 
\end{align}
with $ \Vec{r} \in \mathbb{R}^{2}$ and
\begin{align*}
    \Omega =\begin{pmatrix}
    0& 1\\
    -1 &0 
    \end{pmatrix},
\end{align*}
being the symplectic form. Note that the formalism of density operators is completely equivalent to the here presented formalism using characteristic functions.
Beyond unitary deterministic processes, these Gaussian maps may also include non-unitary maps representing noise or processes where ancillary modes are measured. In the latter case, however, feed-forward is then assumed to take place, to restore determinism. 

The action of a general Gaussian CPTP-map $\Phi$  on the characteristic function can then be written as~\cite{De-Palma:2015vn}
\begin{align}
\label{eq:gen_Gausssian_map}
    \chi_\rho(\Vec{r})\rightarrow& \chi_{\Phi(\rho)}(\Vec{r}) = e^{-\frac{1}{4}\Vec{r}^T\Omega^T Y \Omega \Vec{r} + i \Vec{l}^T \Omega \Vec{r}}\chi_\rho (\Omega^T X^T\Omega \Vec{r}),
\end{align}
with $X$,$Y$ being $2\times 2$ real matrices, $\Vec{l}$ being a $2$-dimensional real vector, $Y$ being symmetric and fulfilling the following positive semi-definite matrix constraint
\begin{align}
\label{eq:gen_Gausssian_map_ineq}
    Y & \pm i(\Omega - X\Omega X^T)\geq 0 \;.
\end{align}
Notice that Eq.~(\ref{eq:gen_Gausssian_map_ineq}) implies that $Y$ has to be a positive semi-definite matrix. The requirement for positive semi-definiteness needs to hold for both signs, since transposition does not influence the positive (semi-) definiteness of a matrix. 
Symplectic transformations are special cases of the protocols introduced in Eq.~(\ref{eq:gen_Gausssian_map}) and correspond to a class of unitary operations for which the noise matrix $Y$ and the displacement vector $\Vec{l}$ are set to zero, whereas $X \in Sp_{2,\mathbb{R}}$ is a symplectic matrix~\cite{Serafini:2017uz}.

A standard measure of closeness or similarity of quantum states is the fidelity~\cite{Jozsa:1994us}
\begin{align}
\label{eq:fidelity}
    \mathcal{F}(\hat{\rho}_1,\hat{\rho}_2) = \qty( \Tr{\sqrt{ \sqrt{\hat{\rho}_1}\hat{\rho}_2 \sqrt{\hat{\rho}_1}  }})^2.
\end{align} 
For a pure state, this expression can be simplified to
\begin{align}
\label{eq:fidelity-pure}
    \mathcal{F}(\hat{\rho},\ket{\Psi_{\textrm{target}}}\bra{\Psi_{\textrm{target}}}) = \langle \Psi_{\textrm{target}} |\hat{\rho}|\Psi_{\textrm{target}} \rangle.
    \end{align}

The conversion protocols we are investigating only feature a single-mode pure state as input and a single-mode pure state as the target. Note however that depending on the Gaussian CPTP-map that the input state is acted on by, the output could be a mixed state. 

In order to find the Gaussian CPTP-map that best approximates the target for a given input state, we numerically optimize the matrices $X,Y$ and the vector $\vec{l}$ with the cost function being the fidelity, which we aim at maximizing. We can rewrite the fidelity for the characteristic function for the input and the target state as
\begin{align}
    \mathcal{F}(\hat \rho,\hat \rho_{\text{target}}) &= 
    \bra{\Psi_\text{target}} \hat \rho \ket{\Psi_\text{target}} \nonumber  \\
    &=\frac{1}{4 \pi} \int d\Vec{r} \;\chi_{\hat \rho}(\Vec{r})\; \chi_{\hat \rho_{\text{target}}}(-\Vec{r}),
\end{align}
where $\hat \rho_{\text{target}} = \ket{\Psi_\text{target}} \bra{\Psi_\text{target}}$. 

What constitutes as a good fidelity depends on the usage.  However, for the case of quantum computation with encoded qubits, fidelities above 95\% are usually regarded as above threshold, namely correctable via code concatenation~\cite{Douce:2019tu}.
In summary, our conversion protocol works as follows: given the characteristic function of the input state, we transform the input characteristic function according to the Gaussian CPTP-map in Eq.~(\ref{eq:gen_Gausssian_map}). We then maximizee the fidelity between the transformed state and the target state by optimizing $X$, $Y$ and $\Vec{l}$, while still fulfilling Eq.~(\ref{eq:gen_Gausssian_map_ineq}).

\subsection{States and codes of interest}
\label{sec:states_and_codes}

\begin{figure*}[bt!]
	\includegraphics[width=\textwidth]{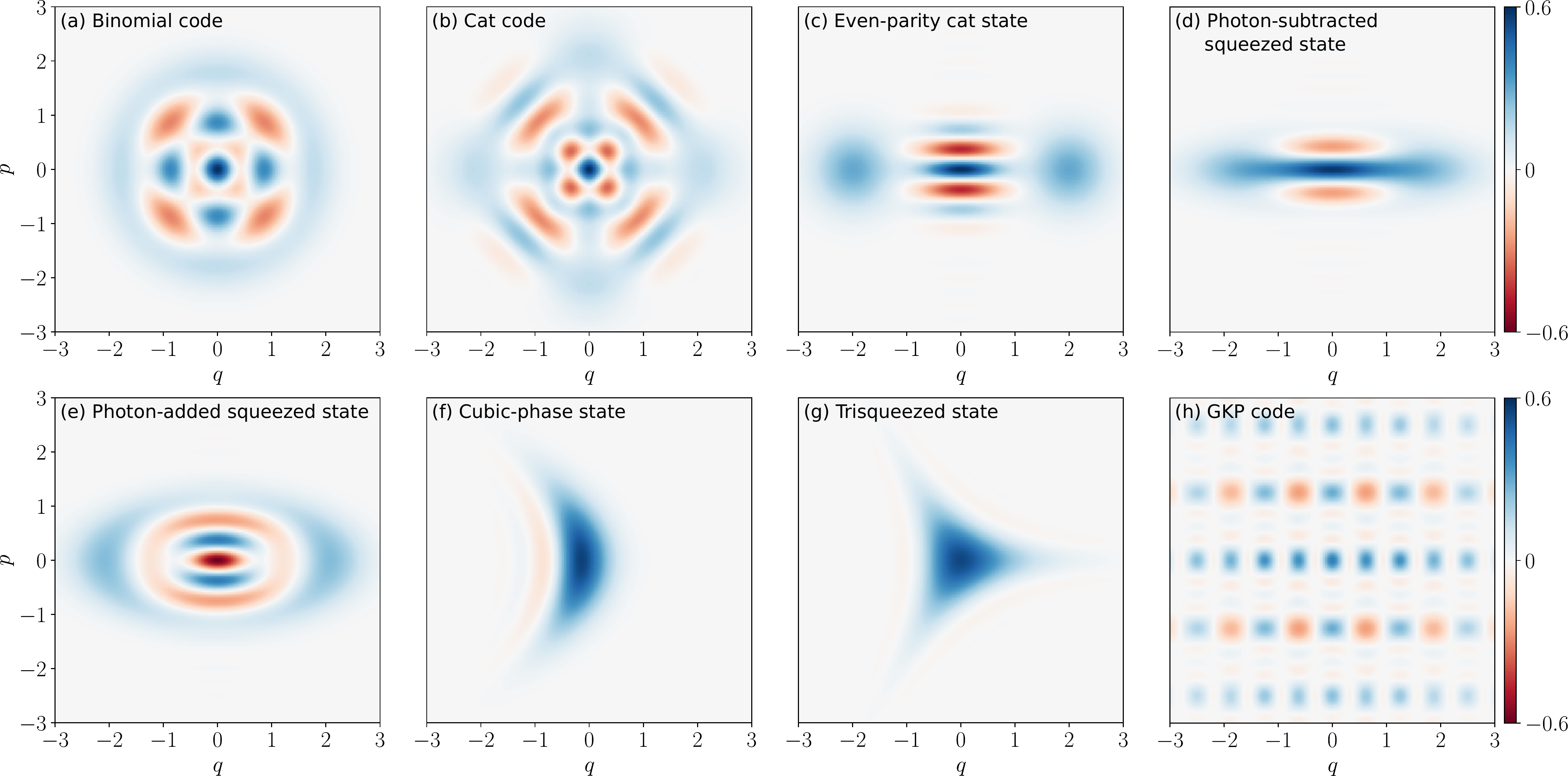}
	\caption{Wigner plots of the types of states and codes used in this paper. The color bar indicates the Wigner positivity (blue) and negativity (red), in normalized units. (a) Binomial code with rotation symmetry $N=2$, truncation $K=3$ and logical encoding $\mu=0$. (b) Cat code with displacement $\alpha=2$, $\mu=0$, $N=2$ and $r=\phi=0$. (c) Even-parity cat state with displacement $\alpha=2$. (d) Photon-subtracted squeezed state with $L=-2$ photons and squeezing $\xi = 5$~dB. (e) Photon-added squeezed state with $L=+3$ photons and squeezing $\xi =3$~dB. (f) Cubic-phase state with cubicity $c=0.551$ and squeezing $-5$~dB. (g) Trisqueezed state (three-photon squeezed state), with triplicity $t=0.1$. (h) GKP code with squeezing $\Delta = 14$~dB, and $\mu=0$ }
	\label{fig:wigner_plot_collage}
\end{figure*}

In this work we investigate conversion between different bosonic codes as well as other known resource states. 
A bosonic code entails the encoding of information in a subspace of the infinite dimensional Hilbert space.
We restrict ourselves to codes that encode qubits and we denote the computational basis states as $\ket{\mu}$ with $\mu \in \{0,1\}$. Figure~\ref{fig:wigner_plot_collage} shows a collage with examples of the codes and states considered.

\subsubsection{Rotationally symmetric bosonic (RSB) codes}
A way to fault-tolerantly encode quantum information into bosonic systems consists of using rotation-symmetric codes~\cite{Grimsmo:2020wa}. RSB codes are designed to protect against photon loss, photon gain and dephasing errors.
These codes are characterized by the order $N$ of rotation symmetry and normalized primitive states $\ket{\Theta}$.
An order $N$-symmetric rotation code has the logical $Z$ operator
\begin{align}
\hat{Z}_N = e^{i(\pi/N)\hat{n}}.
\end{align}
The code-words, i.e. the basis states which encode the 0 and 1 logical information, are defined as
\begin{align}
\label{eq:rot_code}
    \ket{\mu_{\text{Rot}}^{N}}=\frac{1}{\sqrt{\mathcal{N}}}\sum_{m=0}^{2N-1}(-1)^{\mu\cdot m}e^{i\frac{m\pi}{N}\hat{n}}\ket{\Theta}.
\end{align}
The primitive state $\ket{\Theta}$ has to have non-vanishing support on some even and odd Fock numbers.
To the class of rotationally symmetric bosonic codes belong both cat codes and binomial codes.

For the case of cat-codes $\ket{\mu_{\text{cat}}^{N,\alpha}}$, the primitive state that one considers are coherent states, corresponding to the squeezed-coherent states introduced in Eq.(\ref{eq:squeezed-coherent}) in the case of zero phase $\phi=0$ and no squeezing $r=0$
\begin{align}
\label{eq:cat}
    \ket{\Theta_{\text{cat}}} = \ket{\alpha, r=0,\phi=0} = \ket{\alpha}.
\end{align}

Note that in this manuscript we will use the term ``cat states'' to indicate the code-words of a cat code with rotational symmetry of $N=1$, see~\cite{Grimsmo:2020wa}. In particular, the code-words corresponding to $\mu = 0$ yields the even-parity cat state $\propto \ket{\alpha} + \ket{-\alpha}$ while $\mu = 1$ yields the even parity cat state $\propto \ket{\alpha} - \ket{-\alpha}$.
Cat codes with $\alpha \simeq \sqrt{2}$ and $N=2$ have been observed in experiments~\cite{rosenblum2018fault,rosenblum2018cnot, kudra2021robust}. Therefore we will consider here parameters in the range of  $\alpha \in[1 ,3.5]$ and $  N \in \{1, \dots, 6\}$.  

Binomial codes are easier to define in the conjugate basis, where they are expressed as
\begin{align}
    \ket{+_{\text{bin}}^{N,K}} &=  \sum_{k=0}^K \sqrt{\frac{1}{2^{K}}\binom{K}{k}}\ket{kN}, \nonumber \\
     \ket{-_{\text{bin}}^{N,K}} &= \sum_{k=0}^K (-1)^k\sqrt{\frac{1}{2^{K}}\binom{K}{k}}\ket{kN}.
\end{align}
Binomial codes with $N=2$ and $K=2$ have been demonstrated experimentally~\cite{hu2019quantum, kudra2021robust}. In order to include future development, we have chosen $K \in \{2,\dots,6\}$ and $ N \in \{1,\dots,6\}$ as possible parameters for binomial states.
Figures~\ref{fig:wigner_plot_collage} (a) and (b) show Wigner plots of binomial and cat codes, respectively, and (c) shows an even-parity cat state.

\subsubsection{Photon-added and -subtracted squeezed states}
A relevant family of non-Gaussian states that have been experimentally implemented with optical technology are the photon-added and photon-subtracted squeezed states~\cite{lvovsky2020production}. In principle all bosonic quantum states can be created by combining photon addition~\cite{dakna1999erratum} or subtractions~\cite{fiuravsek2005conditional} with linear operations. An example of an important application of photon subtraction is the generation of kitten states (cat states with small amplitude, $\alpha \lesssim 1$)~\cite{lvovsky2020production}, by matching the first two non-vanishing coefficients in Fock basis. 
We define the $L$-photon-added or -subtracted squeezed (PASS) state as ($L\in \mathbb{Z}$)
\begin{equation}
    \ket{{\rm PASS}_L}=\begin{cases}
    \frac{1}{\mathcal{N}}\; \hat{a}^\abs{L} \ket{\alpha, \xi,\phi}, & \text{if $L<0$},\\
    \frac{1}{\mathcal{N'}}\; (\hat{a}^\dagger)^\abs{L} \ket{\alpha, \xi ,\phi}, & \text{if $L>0$},
  \end{cases}
  \label{pass}
\end{equation}

where $\mathcal{N}$ and $\mathcal{N'}$ are normalizing constants. 
Photon-added and photon-subtracted squeezed states are widely used especially for the generation of cat states~\cite{lvovsky2020production}. We considered up to 5 additions and subtractions, with squeezing between $\xi \in[0.1, 1.5]$, corresponding to the range $0.86\textrm{dB}-13 \textrm{dB}$. 
Figures~\ref{fig:wigner_plot_collage} (d) and (e) show Wigner plots of photon-subtracted and photon-added states, respectively.

\subsubsection{Cubic-phase state}
One of the most prominent non-Gaussian states is the cubic-phase state~\cite{Gottesman:2001vk}, shown in Fig.~\ref{fig:wigner_plot_collage} (f). This state can be used to promote purely Gaussian operations to universality~\cite{Lloyd:1999vz, Gu:2009vf}, as well as implementing the crucial non-Clifford T gate for GKP codes~\cite{Gottesman:2001vk}.
The cubic-phase state is defined as 
\begin{equation}
\label{eq:target_state}
    \ket{c}  
   = e^{i c \hat{q}^3} \hat S(\xi) \ket{0},
\end{equation}
where we refer to the parameter $c$ as the \emph{cubicity}.
Due to its fundamental role in quantum computation over continuous variables, various theoretical proposals have been put forward to generate such a state \cite{PhysRevA.84.053802, PhysRevA.88.053816, PhysRevA.91.032321, PhysRevA.93.022301, PhysRevA.97.022329, Arzani:2017ty, Brunelli:2018uq, brunelli2019linear, Sabapathy:2019wu, Yanagimoto:2019ui, Hillmann:2020uf, yu2020,  houhou2022unconditional}, and recently a cubic-phase state was implemented experimentally in microvawe cavities~\cite{kudra2021robust}. 
To chose relevant parameters, we use the Wigner logarithmic negativity~\cite{Albarelli:2018uu, Takagi:2018ul} as a guide, such that the negativity of our target cubic-phase state is comparable to the one of the other states investigated in this work.
The Wigner logarithmic negativity is defined as
\begin{align}
\label{eq:wigner_log_negativity}
M(\hat{\rho}) = \textrm{log}\left(\int d\Vec{r} |W_{\hat{\rho}}(\Vec{r})|\right),
\end{align}
where $W_{\hat{\rho}}(\Vec{r})$ is the Wigner function of the state $\hat{\rho}$, and the integral runs over the whole phase space. This analysis allows us to identify the range $ c \in [0.05,0.3]$ with -5 to 9 dB squeezing for the cubic-phase state to match the Wigner logarithmic negativity of the other states studied in this work.
\subsubsection{Trisqueezed State}
Another non-Gaussian resource state that has been experimentally implemented recently in a microwave architecture~\cite{Chang:2020ts}  is the trisqueezed state~\cite{Braunstein:1987vo, Banaszek:1997ta}, shown in Fig.~\ref{fig:wigner_plot_collage} (g).
The trisqueezed state is defined as 
\begin{align}
    \label{eq:input_state}
  \ket{t} =  e^{i (t^* \hat{a}^3 + t \hat{a}^{\dagger 3})}  \ket{0},
\end{align}
and we refer to the parameter $t$ as its \emph{triplicity}. 
Relevant parameters for the  trisqueezed state are hard to define, since 
this state was only implemented as a steady state. We use again the Wigner logarithmic negativity as above to guide our parameter choice, and we hence limit the range for the triplicity of the trisqueezed state to $t \in[0.1, 0.15]$.  In Ref.~\cite{yu2020}, a reliable Gaussian conversion protocol converting the trisqueezed state onto the cubic-phase state has been identified.

\subsubsection{GKP Code}

An important code in the context of CV quantum computation is the GKP code~\cite{Gottesman:2001aa}, which displays translational symmetry. Thanks to this symmetry, this code was originally designed to protect against small shifts of the quadratures $\hat{q},\hat{p}$.
The code-words are defined as 
\begin{align}
\label{eq:gkp}
    \ket{\mu_{\text{GKP}}^{\text{ideal}}} &\propto \sum_{n \in \mathds{Z}}\ket{\sqrt{\pi}(2n\mu)}_{\hat{q}},
\end{align}
where the index $\hat{q}$ denotes the position eigenbasis. If not otherwise stated, the remaining states are written in Fock basis. The ideal GKP states in Eq.~(\ref{eq:gkp}) are non-normalizable and associated to infinite energy, thus they are not proper quantum states.
To define physical GKP states with finite energy, we consider finitely squeezed GKP states~\cite{Gottesman:2001aa, Albert:2018ui}
\begin{align}
    \ket{\mu_{\text{GKP}}^{\Delta} } \propto \sum_{n \in \mathds{Z}} e^{-\frac{\pi}{2} \Delta^2(2n+\mu)^2   }    \hat{D}\qty(\sqrt{\frac{\pi}{2}}(2n+\mu) ) \hat{S}\qty(-\ln\Delta) \ket{0}, 
\end{align}
where the real parameter $\Delta$ is associated to the squeezing degree. GKP states with about 7 dB squeezing have been implemented in experiments~\cite{Fluhmann:2019aa, Campagne-Ibarcq:2020tq}. 
To encompass experimental improvements, we have chosen 5 to 12 dB squeezing for the GKP states. Figure~\ref{fig:wigner_plot_collage} (h) shows a Wigner plot of a GKP state.

\

\section{Results}
\label{sec:result}
In this section we present various conversions using the CPTP-map defined in Eq.~(\ref{eq:gen_Gausssian_map}).  Let us stress here that, using our numerical tools (see App.~\ref{app:numerics}), we were able to address systematically and exhaustively a variety of conversions and that, in the following, we are going to present only the cases that we deem more relevant.  
For simplicity, when targeting bosonic codes, we chose the code-word corresponding to $\mu=0$.

\subsection{Binomial and cat codes}

\begin{figure*}[bt!]
\includegraphics[width=\textwidth]{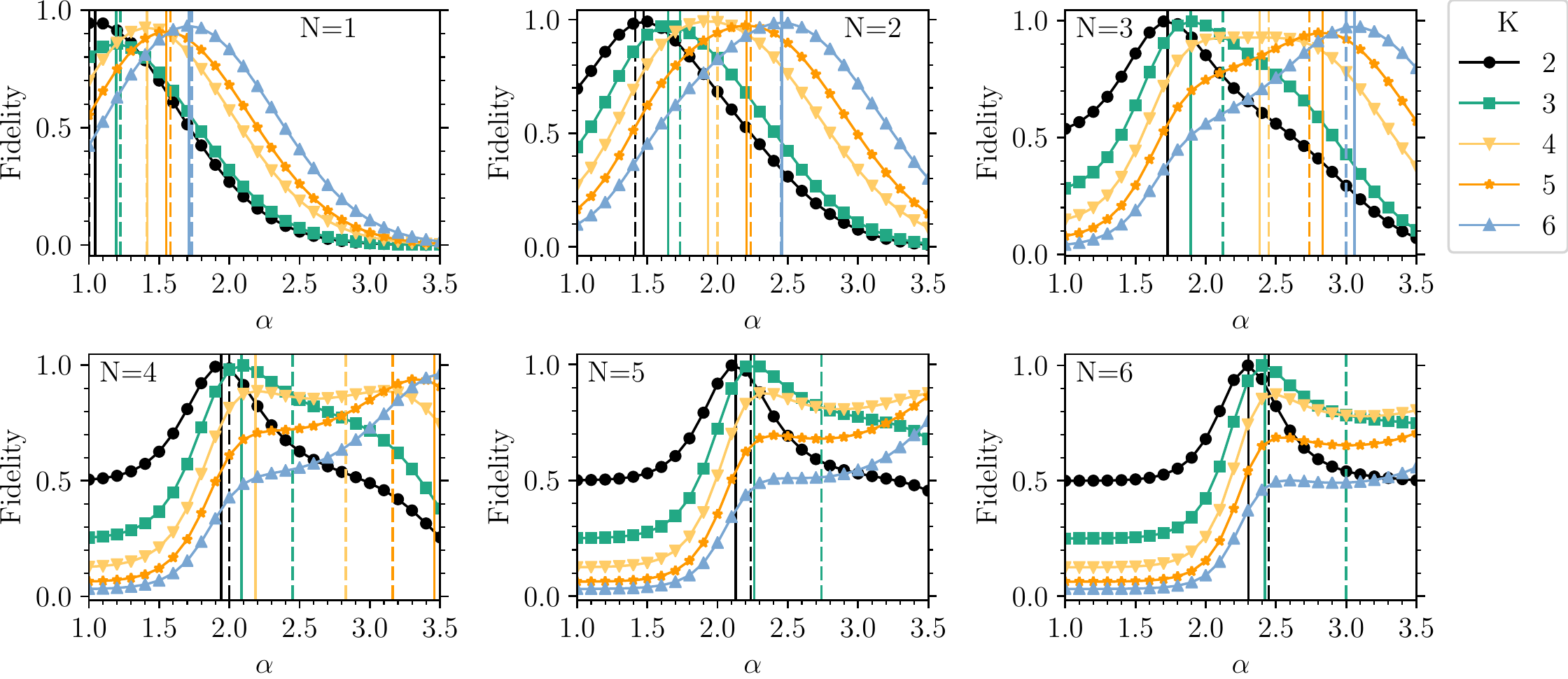}
	\caption{Fidelity between cat and binomial codes as function of the amplitude $\alpha$ of the cat code  for different truncations $K$ of the binomial code. $N$ gives the rotation symmetry of the investigated code-words and is always the same for both. The dotted vertical lines give the value of $\alpha_{\textrm{iso}}$ for which both code-words have the same energy, whereas the solid vertical lines show the value of $\alpha$ corresponding to the maximum fidelity. }
	\label{fig:cat_code_to_binomial_code}
\end{figure*}

\begin{figure}[h!]
	\includegraphics[width=\columnwidth]{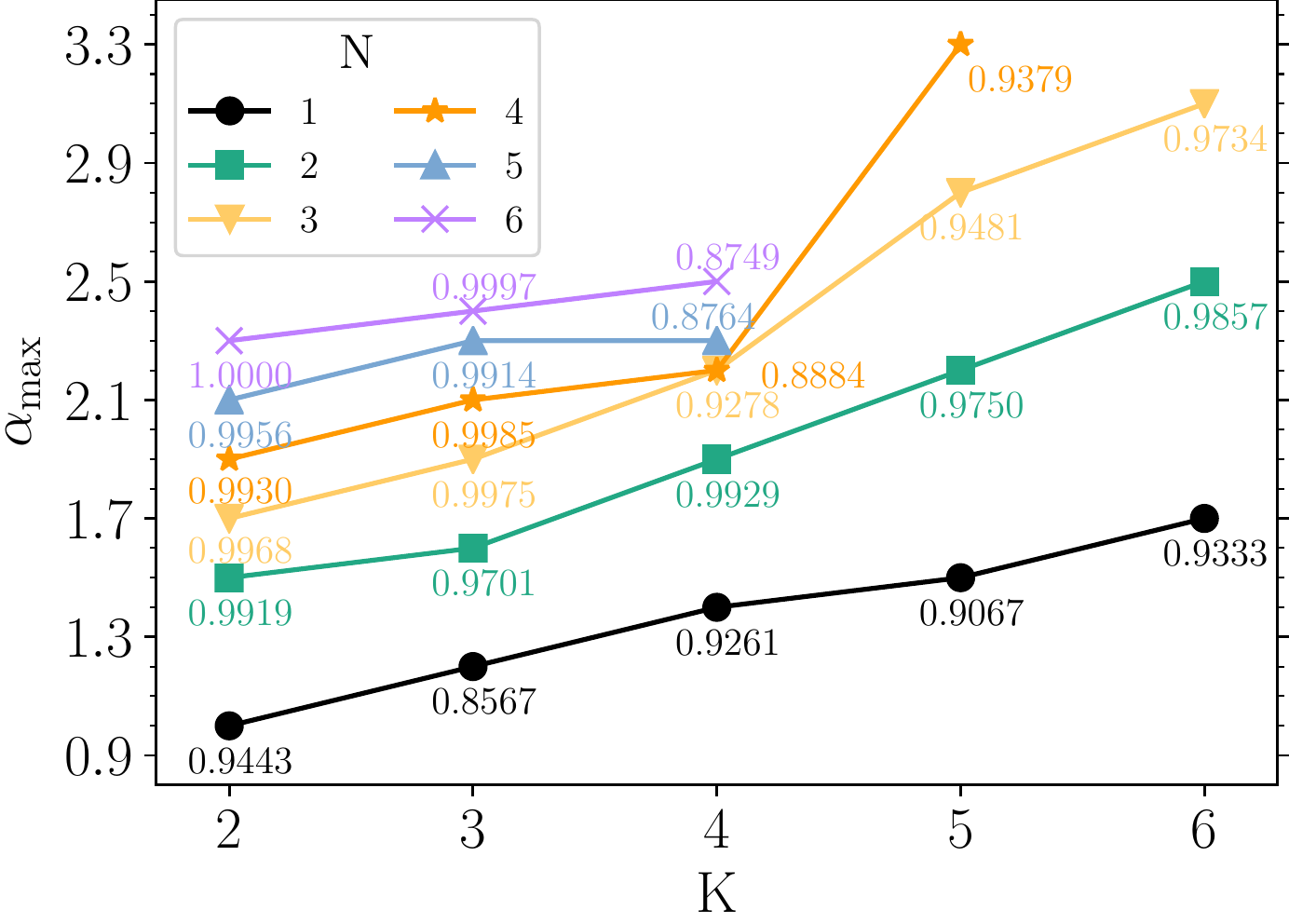}
	\includegraphics[width=\columnwidth]{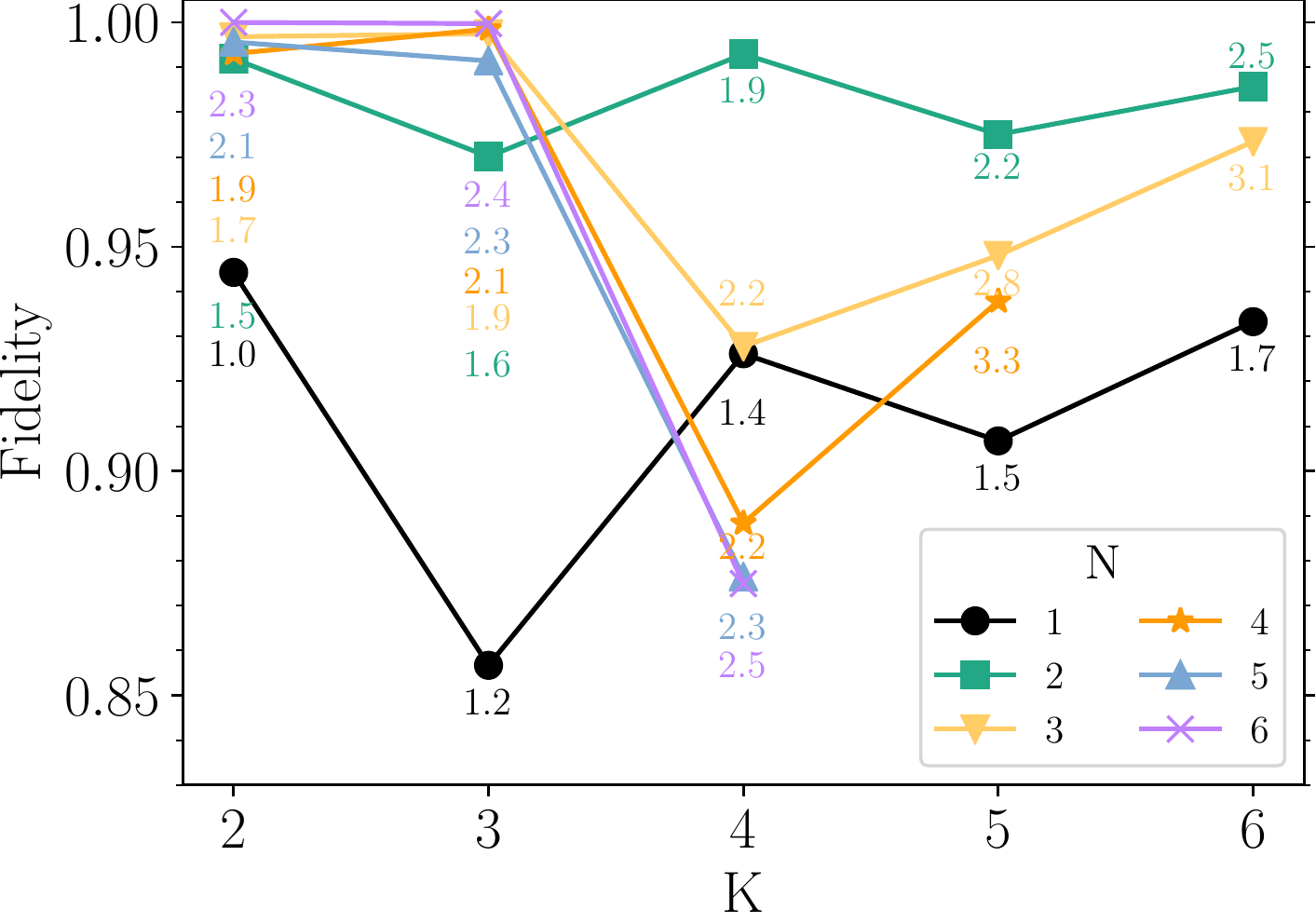}
	\caption{Values of the amplitude $\alpha_{\textrm{max}}$ for the respective cat code at maximum fidelity (top panel) as well as maximum fidelity (bottom panel) as a function of the truncation $K$ of the binomial code for various rotation symmetries $N$. The values next to the markers in the first panel are the values of maximal fidelity, while they are the values of $\alpha_{\textrm{max}}$ in the second panel.}
	\label{fig:cat_code_to_binomial_code_max}
\end{figure}

In this section, we focus on binomial and cat codes. As previously mentioned, these are the most studied instances of rotationally symmetric bosonic codes, and they both have been introduced in the context of error correction to counteract the detrimental effect of losses. The implementation of such codes is currently under intense experimental efforts, and has led to performances beyond the break-even point for quantum error correction~\cite{Ofek:2016wb}. Similarities between these codes are to be expected, given their common symmetric and error correction properties. In fact, for a given rotational symmetry $N$, it is clear that they coincide in the limit of high energy --- namely, in the limit of large truncation ($K \to \infty$) and large displacement ($\alpha \to \infty$)~\cite{Grimsmo:2020wa}. In particular, as noted in Refs.~\cite{Michael:2016uu, Albert:2018ui}, the Fock-state distributions of the binomial and cat codes are binomial and Poissonian, respectively, and they become indistiguishable in the high-energy limit. Besides this asymptotic equivalence, no systematic relation between the two codes is known for the more practically relevant case of finite energy.

In the following we will identify, for a given binomial code, whether a cat code exists such that the two can be considered approximately equivalent \footnote{Notice that the viceversa cannot hold in general, given that cat codes are characterised by continuous parameters whereas binomial codes are characterised by a discrete one. Hereafter, when referring to \textit{equivalence} between cat and binomial codes, we mean that the parameters of the former can be tuned to well approximate the latter, but not the viceversa in general.}. Let us stress here that this equivalence between a priori different codes does not require the active implementation of any Gaussian conversion. In other words, our systematic numerical approach enables us to identify a direct connection between these two codes, with no need of further manipulations of the code states. In addition, as we will see below, such a connection is not immediately intuitive since it cannot be identified by simply selecting isoenergetic code states.

Our numerical findings are illustrated in Fig.~\ref{fig:cat_code_to_binomial_code}. Each of the panels corresponds to a different rotational symmetry $N$, and each curve corresponds to a truncation $K$ of the binomial code. For a given binomial code, we have considered its zero-logical state $\ket{0_\text{bin}^{N,K}}$ and systematically calculated its fidelity with a zero-logical state of a cat code $\ket{0_\text{cat}^{N,\alpha}}$ for different values of $\alpha$. 

Let us focus first on the case with $N=2$. It is clear from Fig.~\ref{fig:cat_code_to_binomial_code} that for any $K$ there exists a value of $\alpha$ such that the fidelity is large (in particular, greater than $0.97$ for the cases under scrutiny). Interestingly, the largest fidelity is in general not achieved for the isoenergetic case. This can be appreciated by considering the fidelity for the case in which the displacement $\alpha$ is set to a value ($\alpha_{\textrm{iso}}$) such that the states $\ket{0_\text{cat}}$ and $\ket{0_\text{bin}}$ have the same energy. The values $\alpha_{\textrm{iso}}$ correspond to the dashed vertical lines in the figure and they do not coincide with the values $\alpha_{\textrm{max}}$ where the maxima of the fidelity curves are located (solid vertical lines in the figure). However, we can see that for larger $K$ the isonergetic states get closer to the maxima, in accordance with the mentioned equivalence of the two codes for high energies. Let us stress here that we exhaustively checked numerically that no Gaussian conversion protocol can improve the fidelity plotted in Fig.~\ref{fig:cat_code_to_binomial_code}, except in regions of low fidelity far from the optimal $\alpha_{\textrm{max}}$.

A detailed inspection of all the results reported in Fig.~\ref{fig:cat_code_to_binomial_code} reveals that the approximate equivalence between binomial and cat codes holds more in general, for larger values of the rotational symmetry $N$, even if not for all the values of the parameters. More specifically, whereas for low and high values of $K$ large fidelities can still be achieved for any $N$, a region of intermediate values of $K$ emerges for which such equivalence does not hold. This behaviour is clearly illustrated in Fig.~\ref{fig:cat_code_to_binomial_code_max} (lower panel), where the maximal fidelity achievable for any pair $(N,K)$ is plotted. Notice that we only show the values of the maximal fidelities for the cases in which it is clear from Fig.~\ref{fig:cat_code_to_binomial_code} that such maxima are in fact achieved, in the range of $\alpha$ considered (\textit{e.g.}, for $N=K=5$ the maximum is not attainded for $\alpha \in [1,3.5]$, and therefore it is not plotted). As said, it is important to stress that, in all the cases of $N$ we considered, the numerical tools developed for our analysis (App.~\ref{app:numerics}) enabled us to show that Gaussian conversion protocols do not help in achieving larger values of the fidelity.

\begin{figure}[h!]
	\includegraphics[width=0.95\columnwidth]{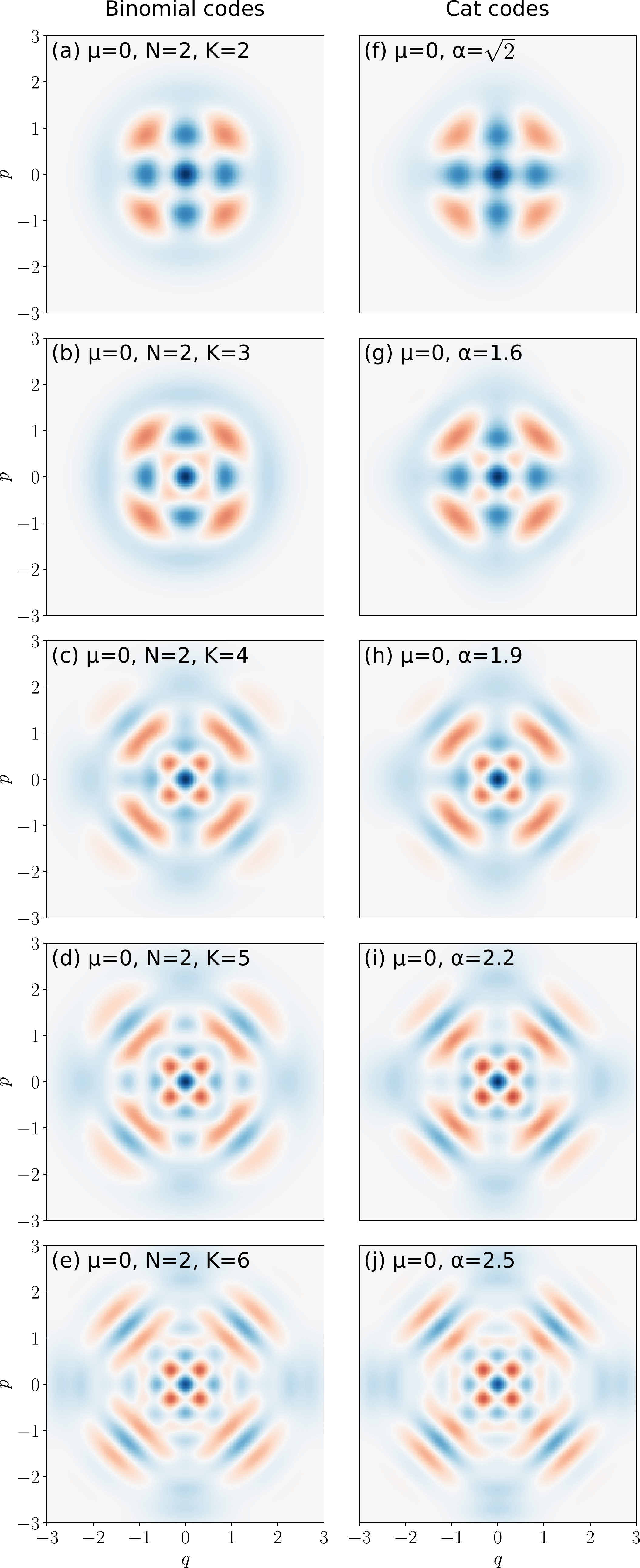}
	\caption{Wigner plots of binomial codes and cat codes, both with rotation symmetry $N=2$, and logical encoding $\mu=0$. The truncation $K$ for the binomial code and the displacement $\alpha$ for the cat code are indicated by the labels. The color scale goes from  $-0.6$ (red) to  $0.6$ (blue). This phase-space representation makes the approximate equivalence between binomial and cat codes becomes apparent.}
	\label{fig:cat_code_to_binomial_code:collage}
\end{figure}

A natural question that stems from the observations above is whether it is possible to find a quantitative relation that determines the approximate correspondence between these two codes. In other words, whether a relation exists (for fixed $N$) between $\alpha_{\textrm{max}}$ and $K$. Our findings are illustrated in Fig.~\ref{fig:cat_code_to_binomial_code_max} (upper panel), where we plot $\alpha_{\textrm{max}}$ versus $K$ for different values of $N$, showing a nearly monotonic increase. This dependence can be intuitively understood by analysing the features of these codes in the phase space. By considering the respective Wigner functions of the two codes, as $K$ increases so does the complexity of the binomial code states (\textit{e.g.}, the number of positive and negative peaks in the phase space increases). Similarly, so does the complexity of the cat code as $\alpha$ increases. Figure~\ref{fig:cat_code_to_binomial_code:collage} shows the Wigner functions for a few different values of $\alpha_{\textrm{max}}$ and $K$, for fixed $N=2$ and $\mu=0$. More in general, from those figures one can appreciate from a phase-space viewpoint the origin of the approximate equivalence of the two codes.

\subsection{Cat states and photon-added and -subtracted squeezed states}

\begin{figure*}[bt!]
\includegraphics[width=\textwidth]{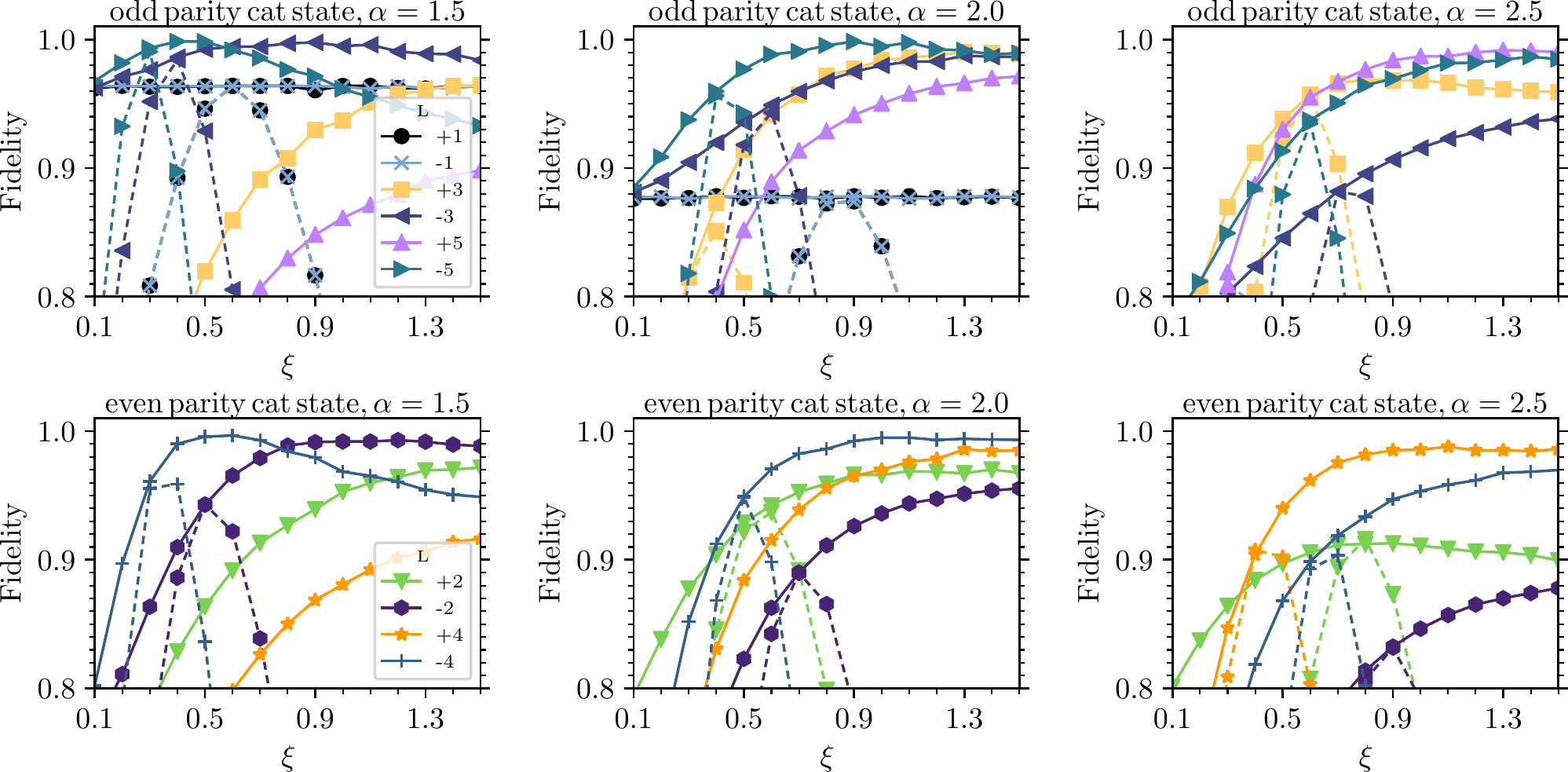}
	\caption{Fidelities of conversion from PASS to cat states as function of the squeezing $\xi$ of the PASS state. Please note that the lowest value of the displacement ($\alpha=1.5$) was chosen such that the cat state no longer looks like a continuous-rotationally symmetric state -- it has well-separated coherence peaks with fringes in-between. The dashed lines are the fidelity between the target cat state and a PASS state with the indicated $L$ and $\xi$. The solid lines are the fidelities after the application of the optimal Gaussian CPTP map that consists of squeezing or anti-squeezing. As can be seen the fidelity with an additional squeezing operation is higher than without the application of a protocol.}
	\label{fig:photon_added_to_cat_state:fixed_target}
\end{figure*}

In this section we focus on Gaussian conversions between PASS and cat states. Close similarities between them have been recognised since the introduction of photon-subtracted squeezed states~\cite{Dakna:1997wk}. In particular, given that the cat state can be produced probabilistically in optical laboratories with relative ease (by combining networks of passive linear optics elements and photon counting), the recognition of such similarities has lead to pivotal experimental achievements such as the generation of kitten states --- namely, cat states with small amplitude $\alpha \lesssim 1$ which display negativities in their Wigner function~\cite{neergaard2006generation, Ourjoumtsev:2007wi, wakui2007photon}. These experimental efforts have been accompanied by intense theoretical investigations, in order to systematically map the relation between these two different classes of states, with the final aim of proposing implementable protocols to generate cat states by photon subtractions and additions~\cite{glancy2008methods, lvovsky2020production}. 

However, this generation technique presents considerable experimental challenges when cat states of larger amplitudes are targeted, since the fidelity between the latter and PASS states is high only for large numbers of photon additions/subtractions which, in turn, imply complex optical networks and low generation probabilities. Here we show that the similarities between PASS and cat states can be significantly boosted by relatively simple Gaussian conversion protocols, even for small numbers of photons additions/subtractions, therefore leading to relevant improvements of the mentioned cat-state generation protocols.

We consider initial PASS states as defined in Eq.~(\ref{pass}) with variable squeezing $\xi$ and $L \in \{-5, \dots,+5\}$. As target we consider cat states with either even or odd parity and variable amplitude $\alpha \in [1.5,2.5]$, as defined in Eq.~(\ref{eq:cat}). For these parameters, the cat states are represented in the phase space by two peaks that are well separated by ``interference fringes'' (a pattern with oscillating Wigner negativity and positivity). This is the regime in which the two components of the cat state become distinguishable enough to enable various applications, including fault-tolerant quantum computation~\cite{lund2008fault}. For any given target, we have optimized the conversion protocol by maximizing the fidelity between the converted state and the target. The maximization is performed numerically over an extensive set of parameters, following the techniques described in Sec.~\ref{sec:cptp_maps}.

The results are presented in Fig.~\ref{fig:photon_added_to_cat_state:fixed_target}, where the fidelity is plotted as a function of squeezing of the input PASS state and the different panels show various target cat states. Each curve corresponds to a unique number $L$ of subtractions or additions, while dashed and solid lines correspond to non-optimized and optimized fidelities, respectively. Specifically, the non-optimized fidelities correspond to the case in which no conversion protocol is applied, and they coincide with the known results available in the literature~\cite{glancy2008methods}. As in the previous section, the curves corresponding to the optimized fidelities are obtained by extensive numerical searches of the optimal conversion algorithm, along the lines discussed in Sec.~\ref{sec:cptp_maps} and App.~\ref{app:numerics}. 

\begin{figure*}[bt!]
    \centering
    \includegraphics[width=\textwidth]{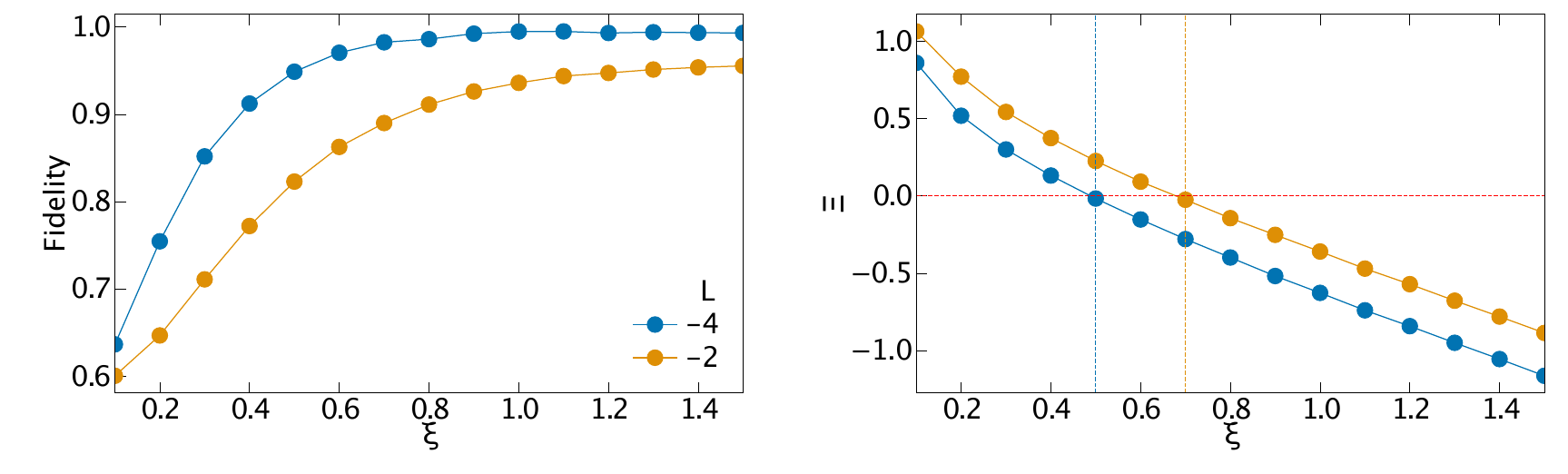}
    \caption{Conversion between a PASS state with  squeezing $\xi$ and $L$ photon subtractions, and an even-parity cat state with $\alpha=2$. The first panel shows the fidelities between the converted PASS state and the cat state, using a Gaussian CPTP, extracted from the central panel in the bottom row of Fig.~\ref{fig:photon_added_to_cat_state:fixed_target}. The Gaussian map is effectively only squeezing with strength $\Xi$. In the second panel we plot the optimal squeezing of $\Xi$ of the map as a function of the input squeezing $\xi$ of the PASS state. The vertical lines show the squeezing for which the PASS state and the cat state have maximal fidelity without application of the map (they correspond to the points where the solid and dashed curves coincide in the central panel in the bottom row of Fig.~\ref{fig:photon_added_to_cat_state:fixed_target}). As can be seen this corresponds to when the Gaussian map does not add any squeezing itself.}
    \label{fig:PASS_2_Cat}
\end{figure*}

In contrast with what was observed in the previous section, we can observe that here, for any set of parameters $(\alpha, L, \xi)$ a Gaussian conversion protocol enables to reach significantly higher values of fidelities with respect to the case when no protocol is applied, except for one point where they coincide. A physical intuition for such an improvement can be gained by considering the Wigner function representing PASS and cat states (see Fig.~\ref{fig:wigner_plot_collage}). As mentioned previously, both states feature two positive peaks along the $q$-axis, separated by fringes. A close inspection of the parameters characterizing the optimized conversion protocols show that they consist mainly in either a squeezing or an anti-squeezing operation (along the $q$-axis) of the PASS state at hand.
The values of the corresponding squeezing parameters for the conversions are plotted in Fig.~\ref{fig:PASS_2_Cat} for $\alpha=2$ in the target cat state for an input PASS state with $L=-2$ and $L=-4$.
Therefore, the action of the conversion is to either increase or decrease the separation of the peaks in the PASS state under consideration, in order to match the separation of the target cat state.

At a more refined level, one can furthermore observe that, as the amplitude $\alpha$ of the target increases, so does the separation between the coherence peaks, as well as the number of fringes. We note that even (odd) parity cat states have a positive (negative) central peak. Similarly, looking at the PASS states, an even (odd) number of photon subtractions/additions also give a positive (negative) central peak. The number of subtractions or additions generally controls the number of fringes between these peaks. Hence, an increased separation and oscillations in the cat state (caused by a higher displacement), has to be matched not only by the right amount of squeezing, but also by the proper amount of photon additions/subtractions, as illustrated in Fig.~\ref{fig:photon_added_to_cat_state:fixed_target}. 

Moreover, notice that most curves in Fig.~\ref{fig:photon_added_to_cat_state:fixed_target} show a monotonic increase in fidelity with squeezing, up to an asymptotic value. This is due to the fact that the conversion protocol can compensate for the squeezing of the PASS state under consideration (by anti-squeezing it, when needed), once the maximal value of fidelity --- which is fundamentally dictated by the value of photon additions/subtractions --- is achieved. 

By way of examples of the improvements attainable with our conversion protocols over known results, let us consider some specific values of the fidelity for the case in which the target is an even-parity cat state with amplitude $\alpha = 2$. This scenario is shown in Fig.~\ref{fig:PASS_2_Cat}. With two photon subtractions at disposal ($L=-2$), the best fidelity achievable without conversion is $F=0.891$, obtained for a PASS state with $L=-2$ and $\xi=0.7$. On the other hand, using our optimized conversion protocol (specifically, given by a squeezing operation of amount $\Xi = - 1.06$), the value of $F=0.95$ can be achieved using a PASS state with $L=-2$ and $\xi=1.3$. Slightly larger fidelities can be obtained for larger $\xi$. Similarly, with four photon subtraction at disposal, the maximal fidelity achievable without conversion is given by $F=0.95$, obtained for a PASS state with $L=-4$ and $\xi=0.5$. By using a conversion protocol this can be improved to $F=0.995$, considering a PASS state with $L=-4$ and $\xi=1$ and enacting on it with an additional squeezing of amount $\Xi = - 1.16$. Namely, an almost perfect conversion can be attained in this case.

Finally, let us notice that our approach share some similarities with the one taken by Menzies and Filip in Ref.~\cite{menzies2009gaussian}. There, a minimal non-Gaussian ``core state'' is identified for any given target state, in particular for a given cat state. Then, the fidelity with the target is optimised with the help of additional Gaussian operations. The latter would correspond to a Gaussian conversion protocol, included in the set of protocols considered here. However, a notable difference with respect to the approach in Ref.~\cite{menzies2009gaussian} is that in our case no ``core state'', specific to the target under consideration is used. Rather we consider a given pair of initial and target states --- PASS and cat states, respectively --- and then identify an optimal Gaussian conversion between the two.

\subsection{Trisqueezed and cubic-phase states}
\begin{figure}[bt!]
	\includegraphics[width=0.9\columnwidth]{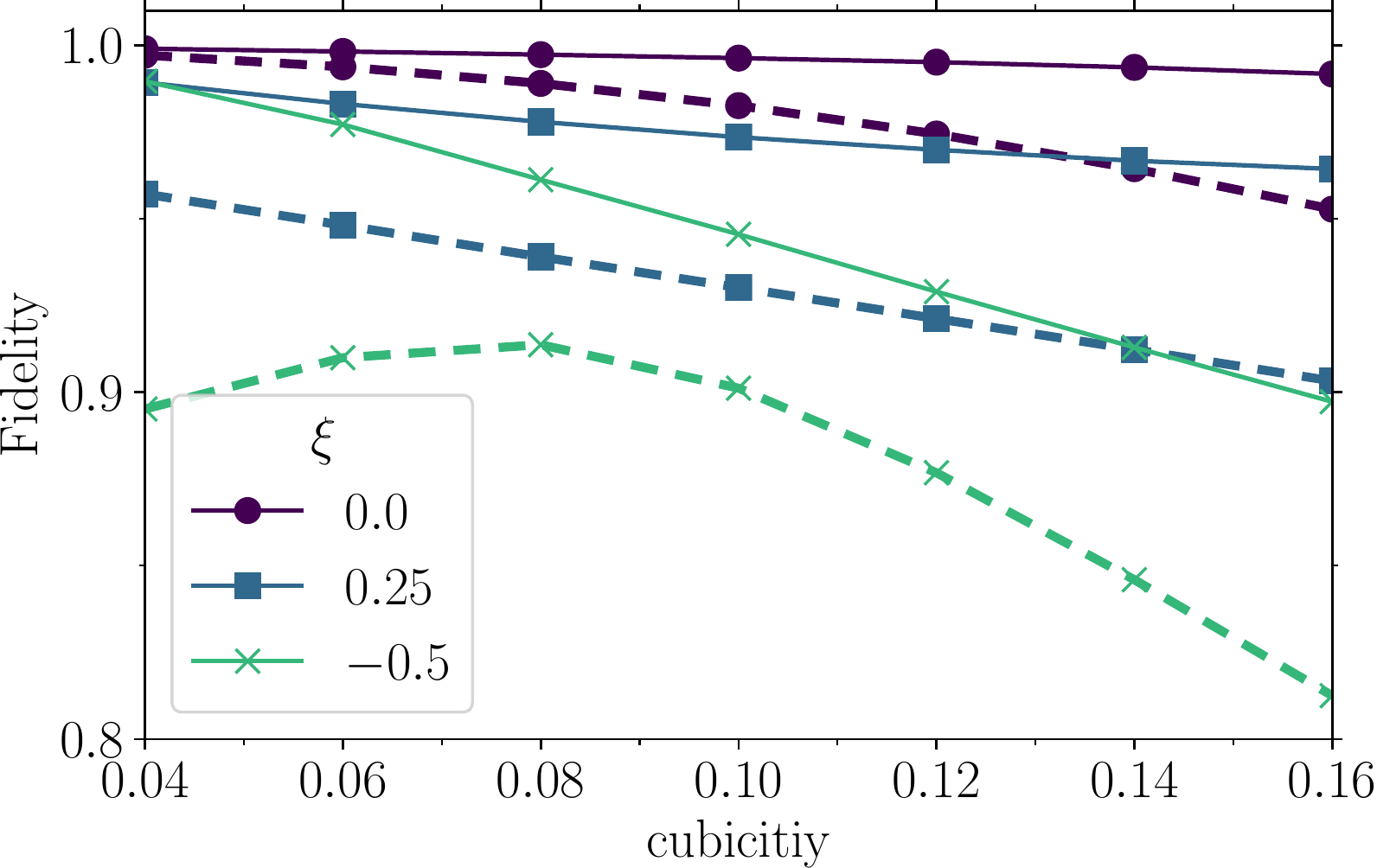}
	\caption{Fidelity for the conversion from trisqueezed to cubic-phase state, as function of the cubicity of the target state. Different markers correspond to different squeezing values $\xi$ of the cubic-phase state, when using an optimal Gaussian CPTP map (solid lines), and no map (dashed lines). The triplicity is such that the Wigner negativity is the same for the input and target state (see values in Tab.~ \ref{tab:unitary_protocol_aec:values}).}
	\label{fig:3photon_to_cubic}
\end{figure}

This section describes conversions between the trisqueezed (3-photon squeezed state) and the cubic-phase state, extending the results presented in Ref.~\cite{yu2020}. We vary the cubicity $c\in [0.04, 0.06, \ldots, 0.16]$ and squeezing $\xi \in [0, 0.25, -0.5]$ extending the parameters from the three fixed conversions studied in Ref.~\cite{yu2020} to a whole range. Following the approach in Ref.~\cite{yu2020}, we choose the triplicities such that the Wigner logarithmic negativity of the input and target states match (see Table~\ref{tab:unitary_protocol_aec:values}). We then run our optimization protocol with the aim of finding the best protocol to convert the chosen trisqueezed state to the target cubic-phase state. We find relatively high fidelities for low cubicities, see Fig.~\ref{fig:3photon_to_cubic}. As the cubicity increases, so does the complexity of the cubic-phase state, until a point where the trisqueezed state simply cannot match all the rich features. Therefore, there is a monotonic decrease in the fidelity. In the more trivial case $\xi=0$, the initial fidelity and optimized fidelity are quite similar, but in all other cases, the protocol gives a significant improvement.
As in Ref.~\cite{yu2020}, the optimal protocol consists of squeezing and small displacements along the $p-$axis.

\begin{table}[!htbp]
\centering
\begin{tabular}{|c|c c c c c c c|}
\hline
cubicity & 0.04 & 0.06 & 0.08 & 0.1 & 0.12 & 0.14 & 0.16  \\
\hline\hline
$t$ ($\xi=0.25$) & 0.048 & 0.063 & 0.075 & 0.084 & 0.0922 & 0.0988 & 0.104\\
$t$ ($\xi=0$) & 0.027 & 0.037 & 0.0464 & 0.0543& 0.061& 0.067  & 0.073\\
$t$ ($\xi=-0.5$) & 0.078 & 0.095 & 0.107 & 0.116 & 0.124 & 0.130 & 0.136\\
\hline
\end{tabular}
\caption{Triplicities ($t$) for Fig.~\ref{fig:3photon_to_cubic}.The triplicities are chosen such the Wigner logarithmic negativity of the input trisqueezed state matches the one of the target cubic-phase state given a certain cubicity and amount of squeezing.
}
\label{tab:unitary_protocol_aec:values}
\end{table}

\subsection{Gaussian no-goes}
\begin{figure}[bt!]
	\includegraphics[width=0.9\columnwidth]{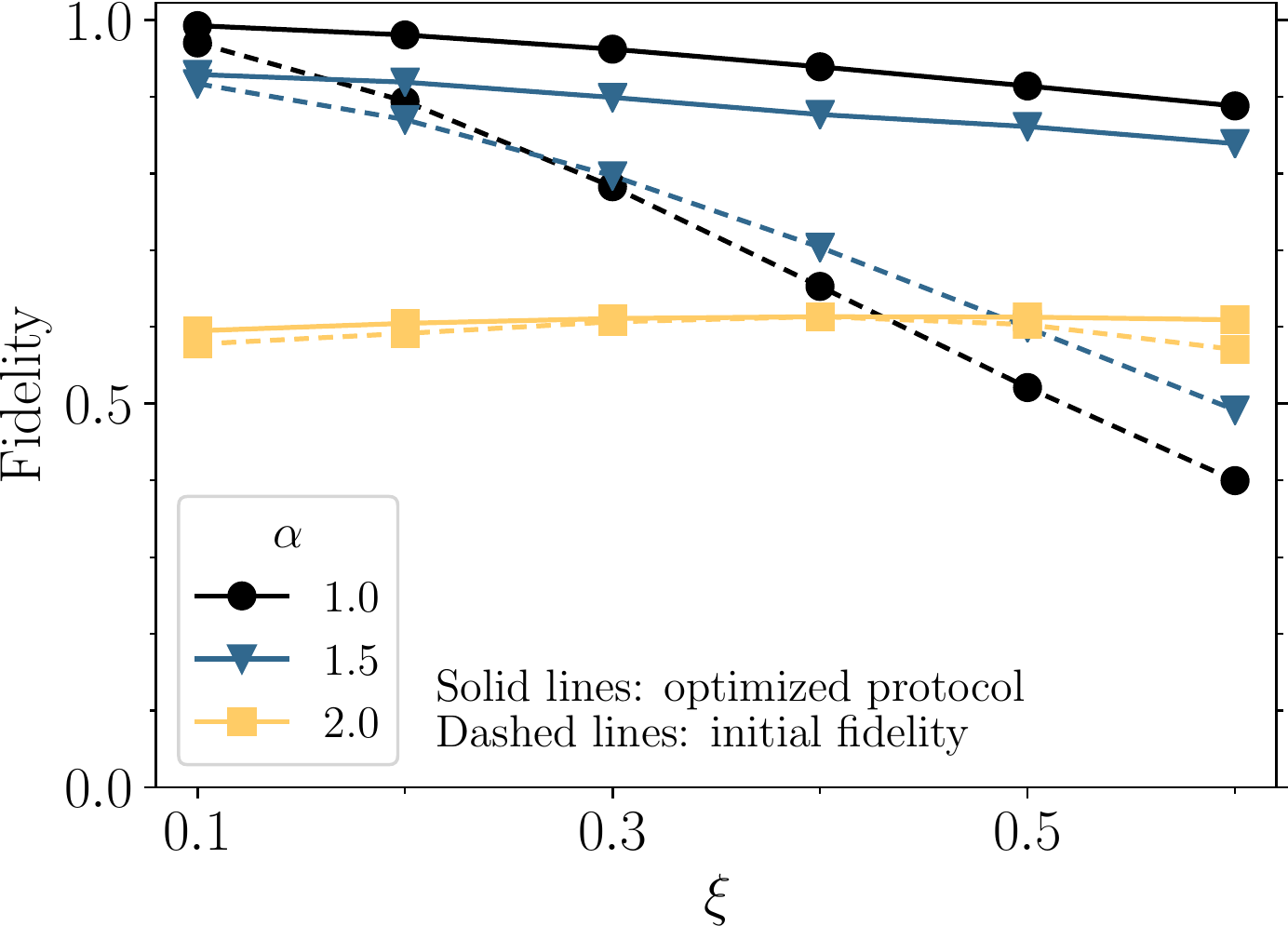}
	\caption{Fidelity for the conversion from a PASS state with $L=2$ added photons to a cat code with rotation symmetry $N=2$, as function of squeezing $\xi$ of the PASS state. The logical encoding of the cat code is $\mu=1$, and the displacement $\alpha$ is indicated by the legend. For low squeezing $\xi$, the states initially look very similar and have a high initial fidelity, but closer inspection in Fig.~\ref{fig:photon_added_squeezed} shows that they have different qualitative features. 
	}
	\label{fig:photon_added_to_cat_code}
\end{figure}

\begin{figure}[bt!]
	\includegraphics[width=\columnwidth]{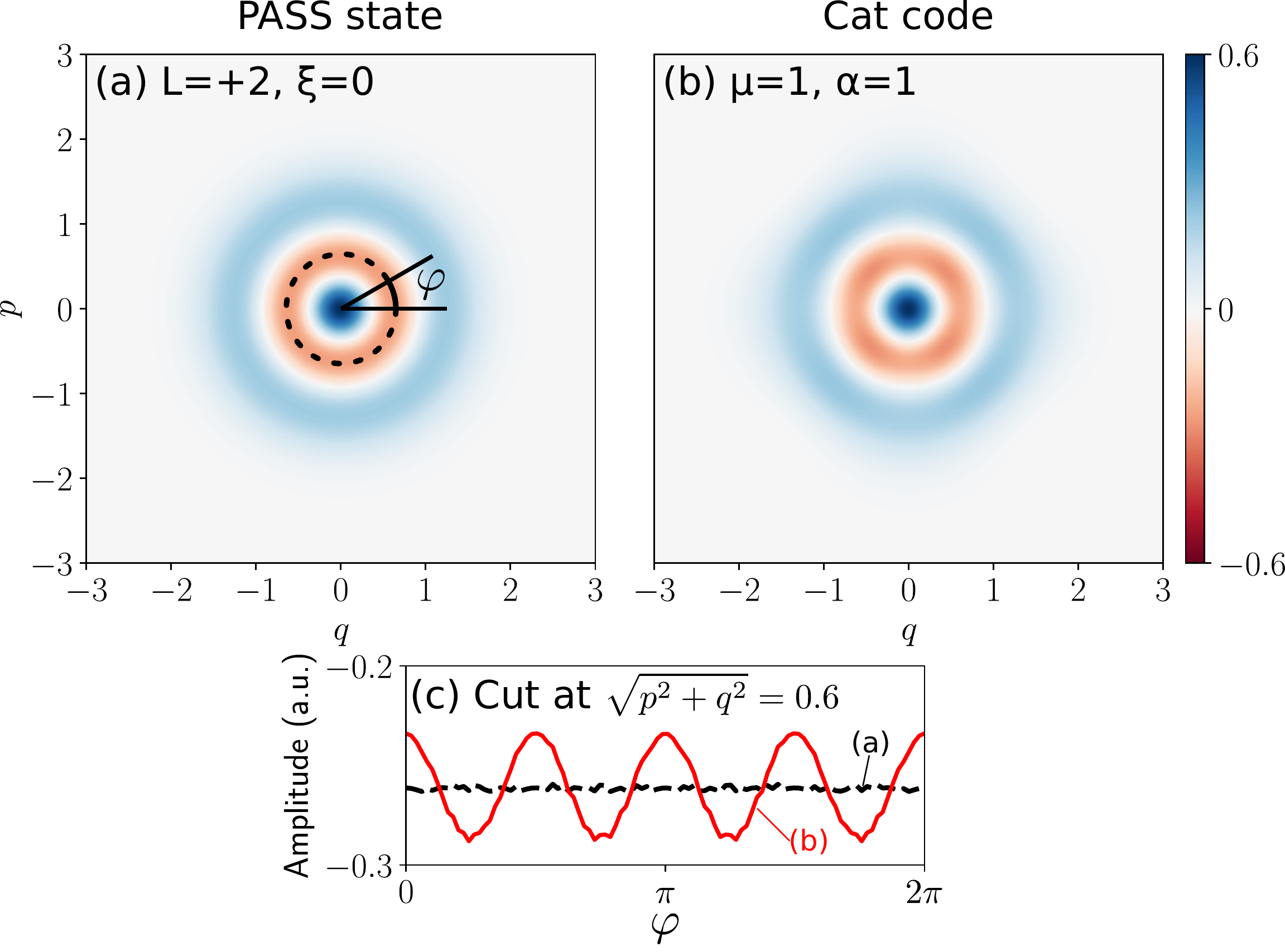}
	\caption{Comparison of rotationally symmetric codes. (a) PASS state with $L=+2$ photons and zero squeezing which is the Fock state $\ket{2}$. (b) Cat code with $\alpha=1$, logical encoding $\mu=1$ and rotation symmetry $N=2$. 
	As the squeezing of the PASS increases, the protocol can still yield a high fidelity by ``undoing'' the squeezing. As the magnitude of $\alpha$ increases significantly, the protocol can also yield a relatively high fidelity, but the cat code breaks the continuous rotational symmetry, displaying a clear four-fold discrete rotational symmetry in phase space. This means that the features are not captured by the protocol, and the conversion is in this sense a no-go. This is illustrated in panel (c), where the amplitude of the Wigner function is plotted along a circle of fixed radius, as indicated by the dashed line in panel (a). The oscillations in the cat code are due to the discrete rotational symmetry, while the photon state is completely rotationally symmetric.}
	\label{fig:photon_added_squeezed}
\end{figure}

\begin{figure}[bt!]
	\includegraphics[width=\columnwidth]{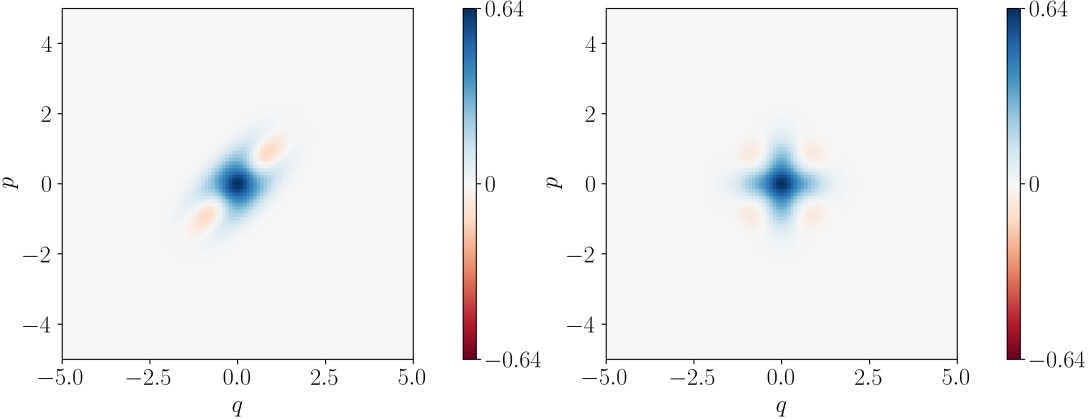}
	\caption{Conversion of a PASS state with $L=-2$ photons and and squeezing $3$~dB to a cat code with displacement $\alpha=1$ and $\mu=0$. The state (left) obtained after applying the conversion protocol to the input state has increased  fidelity from roughly $0.67$ to $0.98$, but misses the Wigner negativity in the top-left and bottom-right corner regions of the target state (right).}
	\label{fig:low_in_hi_out:photon_sub_to_cat_code}
\end{figure}

\begin{figure}[bt!]
	\includegraphics[width=\columnwidth]{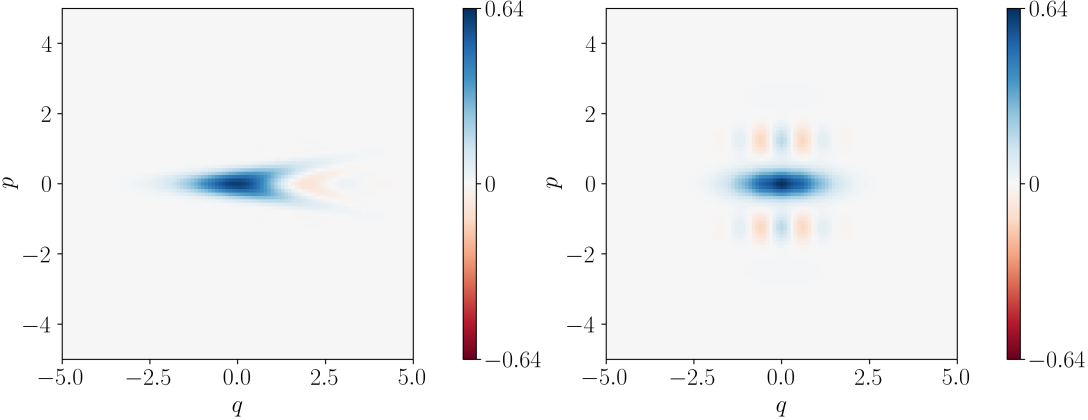}
	\caption{Conversion of a cubic-phase state with cubicity $0.05$ and squeezing $-5$~dB to a GKP state with squeezing noise $5$~dB. The state (left) obtained after applying the conversion protocol to the input state has increased fidelity from roughly $0.55$ to $0.93$, but as can be seen, the qualitative features of the target (right) are captured very poorly.}
	\label{fig:low_in_hi_out:cubic_to_gkp}
\end{figure}

As stated in section Sec.~\ref{sec:states_and_codes}, we have performed an exhaustive and systematic study of state conversions between a variety of states and codes, for wide ranges of parameters. Dealing with such large-scale optimization can lead to pitfalls in the data analysis. In this section, we present conversions that might erroneously be identified as promising, based primarily on the fact that they yield from a low input fidelity to a high output fidelity, and a protocol appearing to do something meaningful. Upon closer inspection, however, these conversions fail to capture the correct qualitative features of the target state. These findings can be framed along the ones of Ref.~\cite{dodonov2011upper,bina2014drawbacks, Mandarino2016}, where it is shown that the fidelity can have a limited predictive power of the relevant features for quantum states. Table~\ref{tab:gaussian_no_go:1} summarizes a few such conversions. Figures~\ref{fig:photon_added_to_cat_code}--\ref{fig:photon_added_squeezed} show the notable example of conversion between PASS states with $L=2$ to cat codes, which has a relatively high fidelity, and at first glance very similar Wigner functions. Under closer inspection, however, the PASS states have continuous rotational symmetry, while the cat codes have a discrete rotational symmetry. Two other notable examples are PASS states $L=-2$ to cat codes, and cubic-phase states to GKP codes, illustrated in Figs.~\ref{fig:low_in_hi_out:photon_sub_to_cat_code} and \ref{fig:low_in_hi_out:cubic_to_gkp}, respectively. Although the fidelities are relatively high, these figures show significant discrepancies in the Wigner functions. In particular, the negativity of the quasi-probability distributions is known to yield an important characterisation of the non-classicality of the state \cite{Wigner:1932tp, banaszek1999testing, spekkens2008negativity, ferraro2012nonclassicality} and indeed the Wigner negativity is necessary \cite{Mari:2012aa, veitch2012negative}, even if not sufficient \cite{Garcia-Alvarez2020}, for quantum computational advantage. As it can be seen in Figs.~\ref{fig:low_in_hi_out:photon_sub_to_cat_code} and \ref{fig:low_in_hi_out:cubic_to_gkp}, the negativity features of the the states under consideration are captured very poorly. These results further emphasize that the fidelity is not the best measure to determine closeness in an operational sense, and motivate the use and discovery of other metrics.

\begin{table}[!htbp]
\centering
\begin{tabular}{| c | c | c | c |}
\hline
    Input code/state & Target code/state & Init.Fidelity & Fidelity  \\
    \hline
    PASS (L=-2, 1 dB) & GKP(6 dB) & 0.8579 & 0.9189 \\
    PASS (L=-2, 3 dB) & GKP(5 dB) & 0.8762 & 0.9361 \\
    PASS (L=-2, 1 dB) & GKP(5 dB) & 0.9319 & 0.9630 \\
    \hline
    PASS (L=-2, 5 dB) & Cat($\alpha$=1, $\mu$=0) & 0.3891 & 0.9346 \\
    PASS (L=-2, 3 dB) & Cat($\alpha$=1, $\mu$=0) & 0.6711 & 0.9791 \\
    PASS (L=-2, 1 dB) & Cat($\alpha$=1, $\mu$=0) & 0.9199 & 0.9701 \\
    \hline
    PASS (L=+2, 5 dB) & Cat($\alpha$=1, $\mu$=1) & 0.4279 & 0.8954 \\
    PASS (L=+2, 3 dB) & Cat($\alpha$=1, $\mu$=1) & 0.7250 & 0.9518 \\
    PASS (L=+2, 1 dB) & Cat($\alpha$=1, $\mu$=1) & 0.9617 & 0.9914 \\
    \hline
    PASS (L=+2, 5 dB) & Cat($\alpha$=$\sqrt{2}$, $\mu$=1) & 0.4976 & 0.8659 \\
    PASS (L=+2, 3 dB) & Cat($\alpha$=$\sqrt{2}$, $\mu$=1) & 0.7547 & 0.9127 \\
    PASS (L=+2, 1 dB) & Cat($\alpha$=$\sqrt{2}$, $\mu$=1) & 0.9323 & 0.9516 \\
    \hline
    Cat($\alpha=1$, $\mu=0$) & Cubic(c=0.05, -5 dB) & 0.8241 & 0.9560 \\
    Cat($\alpha=1$, $\mu=0$) & Cubic(c=0.1, -5 dB) & 0.7382 & 0.9222 \\
    Cat($\alpha=1$, $\mu=0$) & Cubic(c=0.05, -7 dB) & 0.7260 & 0.9226 \\
    \hline
    Cat($\alpha=1$, $\mu=0$) & GKP(5 dB) & 0.8764 & 0.9243 \\
    \hline
    Cubic(c=0.05, -5 dB) & GKP(5 dB) & 0.5485 & 0.9257 \\
    \hline
    trisqueezed ($t=0.1$) & GKP(5 dB) & 0.7741 & 0.9171 \\
    trisqueezed ($t=0.1$) & Cat($\alpha=1$) & 0.9082 & 0.9195 \\
\hline
\end{tabular}
\caption{Examples of conversions where the Gaussian conversion appears to be efficient, but actually fails to capture the correct features (or is uninteresting/trivial for some other reason). }
\label{tab:gaussian_no_go:1}
\end{table}

\section{\label{sec:conclusions}Conclusions}

In this paper we have presented an exhaustive study of single mode Gaussian conversions between non-Gaussian bosonic states, based on  numerical simulations.
Our numerical framework is in the progress of being released as open source, and enables large-scale throughput of conversions between different states and codes, with optimization over wide ranges of parameters. For more information consult appendix~\ref{app:numerics}.

First, we identified an approximate equivalence between cat and binomial code-words. For all the cases we tested, no improvement was possible through the usage of Gaussian CPTP maps.
This correspondence was previously known only in the asymptotical case, while we showed that these codes are approximately equivalent for certain sets of parameters in the finite-energy regime. 
Second, we demonstrated an improvement in the conversion of PASS to cat states by applying a Gaussian CPTP map. The approximate equivalence from PASS to cat states is known since the introduction of PASS states, however we showed that the fidelties can be increased substantially by additional squeezing/antisqueezing.
Furthermore, we showed improvements in the fidelity of the conversion from trisqueezed to cubic-phase states through the application of Gaussian maps for a wide range of parameters, extending the results presented in~\cite{yu2020}.
Finally we presented a few deceptively good conversion that do not reproduce important features, illustrating
that the fidelity alone is not a sufficient measure to estimate successful conversions.

Based on our results, it appears that the Gaussian CPTP map is a quite limited protocol. Consequently, successful conversions such as in Ref.~\cite{yu2020}, stem from our systematic work as  exceptions, rather than the rule. As a potential remedy, we suggest Gaussian CPTP maps with multi-mode distillation, and probabilistic protocols. These protocols are being implemented into our numerical framework, and further studies are underway.

\begin{acknowledgments}

P.H. and O.H. have contributed equally to this work. 
G.F. acknowledges support from the Swedish Research Council (Vetenskapsrådet) through the project grant QuACVA. G.F., O.H. and P.H. acknowledge support from the Knut and Alice Wallenberg Foundation through the Wallenberg Center for Quantum Technology (WACQT).
The computations were enabled by resources provided by the Swedish National Infrastructure for Computing (SNIC) at NSC partially funded by the Swedish Research Council through grant agreement no. 2018-05973. 
\end{acknowledgments}

\appendix

\section{FidelityOptim: a numerical framework for fidelity optimization}
\label{app:numerics}

The fidelity calculation presented in Sec.~\ref{sec:background} represents a challenging computational problem. Indeed, using publicly available tools (such as QuTiP~\cite{Johansson:2012uz,Johansson:2013wy}), it can take several hours to evaluate a single fidelity. It can therefore become unfeasible to solve the optimization problem of finding the best conversion parameters for the Gaussian CPTP map, as this requires mapping out large parameter phase spaces, and sometimes evaluating over thousands of fidelities. Furthermore, the complexity scales rapidly with the number of independent optimization parameters of the protocol. 
Motivated by a need for large-scale optimization, we have developed an efficient and versatile numerical framework, called FidelityOptim, with a powerful backend written in C++ and CUDA\cite{NVIDIA-Corporation:2019tc,Nickolls:2008vi}, and a user-friendly frontend written in Python. The frontend interface removes the need for the user to write source code, making it trivial to set up large-scale conversion optimizations, as well as to post-process and analyze the results. FidelityOptim can compute over hundreds of fidelities per second on a desktop computer, making it possible to explore relatively large phase spaces, and find the optimal conversions. This is made possible by: 1) an efficient implementation of the Gaussian CPTP-map; 2) exploiting parallelism and high-performance computing; 3) using state-of-the-art hardware in the form of Graphics Processing Units, and 4) using efficient optimization strategies.
For comparison and portability reasons, the code also runs relatively fast on Central Processing Units. This makes FidelityOptim useful on anything from consumer laptops and desktops, to powerful nodes in cluster environments.

To ensure that we do not miss optimization parameters where the Gaussian conversion protocol might provide a better fidelity, we use the robust and versatile Particle-Swarm Optimization strategy~\cite{Kennedy:1995th,Yang:2011wv}. With this method, we compare thousands of different parameters in parallel, for each iteration. We typically start by spreading the swarm either randomly or uniformly in the optimzation-parameter phase-space, and let the swarm search the phase space until it converges towards an optimal fidelity. 
For more details on the optimization strategy, see App.~B in Ref.~\cite{yu2020}. 

\bibliography{bib}

\end{document}